\documentstyle[12pt]{article}
\def\gev{{\rm \, Ge\kern-0.125em V}}
\begin{document}
\begin{titlepage}
\pagestyle{empty}
\baselineskip=21pt
\rightline{McGill 96--34}
\rightline{hep-th/9609087}
\rightline{September 1996}
\vskip .2in
\begin{center}
{\large{\bf Stringy Toda Cosmologies}}
\end{center}
\vskip .1in
\begin{center}
Nemanja Kaloper

{\it Department of Physics, McGill University}

{\it Montr\'eal, Qu\'ebec, Canada H3A 2T8}

\vskip .1in

\end{center}
\vskip .5in
\centerline{ {\bf Abstract} }
\baselineskip=18pt
We discuss a particular stringy modular 
cosmology with two axion
fields in seven space-time dimensions, 
decomposable as a time
and two flat three-spaces. The effective 
equations of
motion for the problem are those of the
$SU(3)$ Toda molecule,
and hence are integrable. We write 
down the solutions,
and show that all of them 
are singular. They can be thought of 
as a generalization of the Pre-Big-Bang
cosmology with excited internal 
degrees of freedom, and still 
suffering from the graceful exit problem. 
Some of the solutions 
however show a rather unexpected
property: some of their spatial sections shrink to a 
point in spite of
winding modes wrapped around them.
We also comment how 
more general, anisotropic, solutions, 
with fewer Killing symmetries 
can be obtained with the help of
STU dualities.
\vskip.5in
\centerline{\it Submitted to Physical Review {\bf D}}

\end{titlepage}
\baselineskip=18pt
%%%%%%%%%%%%%%%%%%%%%%%%%%%%%%%%%%%%%%%%%%%%%%%%%%%%%%%%%%%%%%%%%%%%%
{\newcommand{\la}{\mbox{\raisebox{-.6ex}{~$\stackrel{<}{\sim}$~}}}}
{\newcommand{\ga}{\mbox{\raisebox{-.6ex}{~$\stackrel{>}{\sim}$~}}}}
\def\beq{\begin{equation}}
\def\eeq{\end{equation}}
%%%%%%%%%%%%%%%%%%%%%%%%%%%%%%%%%%%%%%%%%%%%%%%%%%%%%%%%%%%%%%%%%%%%%
There has been much interest recently in using 
string theory to address the problems of cosmology, 
including the issue of the initial cosmological singularity, 
symbolizing the
Big Bang in the Standard Cosmological Model 
\cite{BrVa}-\cite{CLW}. Perhaps
the best developed approach to date is the Pre-Big-Bang
cosmology \cite{VG}, which originally strove to relate
the singularity problem with inflation. Briefly, the 
idea rested on the rolling dilaton field, which 
would necessarily evolve toward a region of both strong 
string coupling and high curvatures, resulting
in a phase of accelerated expansion of the Universe in
the string frame, where some (as yet
unknown) dynamical mechanism would kick in to prevent
further increase of the curvature and transform
the evolution into
a more moderate power(ed)-law expansion. This ``graceful exit"
has been shown to be very difficult to attain; we know
that the semiclassical corrections alone are not capable of 
producing it \cite{BV,KMO,EMW}. 
It turns out that in the sub-Planckian 
region of curvatures the influence of the terms leading to 
singular behavior in the classical theory is too strong. 
Figuratively speaking, the impending singularity does all the 
``driving" at the scales close to the Planck mass. As a 
consequence, it has become evident that a Pre-Big-Bang phase
must evolve to the Planckian energies, where we would need
a better understanding of the effects of quantum gravity
to see if a graceful exit can occur there. In practical terms,
however, this means that the Pre-Big-Bang cosmology still 
contains a Planck phase, much like the Standard Model it was
meant to supplant \cite{KK}-\cite{EM}. 

On the other hand, it has been suggested that the
singularity can be removed, or at least redefined, by the
winding mode sources in the 
Early Universe \cite{BrVa}. The winding modes
are expected to contribute significantly to the effective
cosmological dynamics at very high energies, or equivalently,
at small distances. This is because their energy is generically 
inversely proportional to the radius of the compact direction
around which they are wound up, and 
hence it grows rapidly as this
radius diminishes. In fact, the presence of the winding modes 
should prevent a complete collapse of the compact directions to 
the zero radius - the growing energy of the winding modes should
overtake all the other effects and cause the scale factor to
``bounce" away from zero. Indeed, this is known to happen in 
some special cases, as investigated in \cite{BrVa,Ts,CLW}. 
As a matter of fact, the example of reference 
\cite{CLW} is the winding mode in disguise:
the pseudoscalar axion field of 
this solution can be dualized to
its associated three-form, which 
is then seen to be purely magnetic:
it is equal to a constant times 
the volume form of the three-space.
We can then formally compactify 
the three-space to a three-torus,
and think of the magnetic axion 
as a membrane wound around this torus-
i.e. the winding mode source. 
The interpretation of the bounce which
occurs in the scale factor for 
this solution then becomes clearer:
it arises in response to the 
increased resistance of the membrane to
being squeezed down to a point.

Upon observing this fact, one 
might wonder what might happen if two
axions (equivalently, two axionic charges)
are present. Namely, one might 
be able to find cosmological
analogs to solitonic string solutions, 
which are generically nonsingular
because of a convenient redefinition of 
the singularity. Further, the presence
of two axionic charges may lead to the 
stabilization of some of the moduli 
fields, whose singular behavior often 
leads to Planck-scale curvature
singularities. Hence it is interesting 
to study the details of this mechanism,
in order to shed some light on the 
small-scale behavior of winding modes.
Surprisingly, the model we will discuss 
here turns out to be completely
integrable despite its rather complicated 
nonlinear structure. We will
recognize the equations of motion for the 
case of two axionic charges
as those of the $SU(3)$ Toda
molecule, and will therefore be able to 
write down a completely general family 
of classical solutions. The solutions 
in turn do not possess the properties we
have hoped for initially: they all have naked
singularities, either in the past 
(like the $(-)$ branch solutions of the
chargeless case \cite{VG}) or
in the future (like the 
(+) branch solutions \cite{VG}) or both. 
The phase-space dynamics of the model 
is considerably more complex than in the cases without any, 
or with a single axion charge. Depending on the initial
conditions, there are solutions whose complete evolution takes
place in the whole range of coupling strengths, similar to the
situation without charges. There are also 
solutions which evolve entirely in the region of strong
coupling, like in the single charge case 
(such behavior is
expected because of the S-duality relationship of 
string coupling to the chargeless case: 
$g' = c^2 g + d^2 g^{-1}$, where $c,d$ are real numbers;
since $g$ varies on the whole half-line $[0, \infty)$, 
$g'$ is bounded below by $2|cd|$). 
Finally,
some of the solutions evolve entirely in the region of weak 
coupling, and have a rather unexpected behavior in 
that their spatial subspaces oscillate around
zero volume, in spite of 
a winding mode wrapped 
around them. The solutions from this
subfamily will therefore have both
past and future singularities. This is in
sharp contrast to the chargeless 
case, where the presence of a mode
wound around any spatial section 
of nontrivial homotopy ensures that
the section must always have finite 
volume. The effect is a consequence
of a nontrivial interaction between 
the modes, given by the Toda potential,
as a result of which the dynamics is not just
a simple extension of the previous results.
One should keep in mind, however, that our analysis is
classical, and that, in principle, the quantum corrections
could resurrect the bounce expected from the winding modes.
We will finally comment on the use of string dualities 
for generalizing the solutions presented in this paper,
and in particular point out a connection between our
solutions and an anisotropic Bianchi II family
of cosmologies. The novelty of this result is that not only
was isotropy lost but also the algebra of 
the basis forms of the solution has been changed. 
Such changes may turn out to be important
for the considerations of the cosmological
singularity \cite{KK}.

We begin by deriving the simplest 
nontrivial Toda cosmology,
the $SU(3)$ case, as a solution of the 
effective action
describing the graviton multiplet in
seven dimensions. To the lowest order 
in $\alpha'$,
\begin{equation}
\label{sact}
S = \int d^7 x \sqrt{g} e^{- \Phi}
\Bigl\{R + ({\nabla} \Phi)^2 - \frac{1}{12}
H^2_{\mu\nu\lambda} + 2 \Lambda_7 \Bigr\}
\end{equation}
\noindent The fields $\Phi$ and
${H}_{\mu\nu\lambda} = \partial_\mu
{B}_{\nu\lambda} + cyclic~$
$permutations$ are the seven-dimensional
dilaton and three-form axion 
fields respectively.  
The last term, $\Lambda_7$, 
represents the stringy cosmological
constant, which can arise from 
the central charge deficit of the 
target space. We keep it here 
for completeness, although we will
ignore it in the most of what follows.
This action can be thought of 
as a consistent truncation of 
any
superstring theory, containing only
the NS-NS fields in addition to the graviton.
We will look for the solutions of the form
\begin{eqnarray}
\label{dans}
d{s_7}^2 &=& -n(t)^2 dt^2 + a(t)^2 
d\vec x^2 + b(t)^2 d\vec y^2 \nonumber \\
H_7 &=& P d\vec x^3 + Q d\vec y^3 \\
\Phi_7 &=&\Phi(t) \nonumber
\end{eqnarray}
The axionic equations of 
motion can be written in 
form notation as $d^{*}H=dH=0$, and so 
it is not difficult to verify 
that the Bianchi identity
requires that $P$ and $Q$ be 
constant. We could now write down the
equations of motion obtained 
from the action above, substitute the
ans\"atz and look for solutions. 
We choose to work in the action,
however, dimensionally reducing 
the solution by employing the fact that
six coordinates are cyclic, and 
then varying the resulting 1D
action with respect to the 
dynamical degrees of freedom, as well as
the gauge parameter $n$ 
(the lapse function).
The action becomes,
up to a 1D boundary term, 
and after the redefinition of the dilaton
according to $\Phi_7 = \phi + 3\ln a + 3\ln b$,
\begin{equation}
\label{sact1}
S = \int dt e^{- \phi}
\Bigl\{2\Lambda_7 n 
- \frac{n}{2} (\frac{P^2}{a^6} 
+ \frac{Q^2}{b^6}) - \frac{\dot \phi^2}{n}
+ \frac{3 \dot a^2}{na^2} 
+ \frac{3 \dot b^2}{nb^2}\Bigr\}
\end{equation}
Varying this action with 
respect to $a,b,\phi$ and $n$, and then 
choosing the gauge $n=\exp(-\phi)$, we find 
\begin{eqnarray}
\label{eoms}
\dot h &=& \frac{P^2}{2} 
\frac{e^{-2\phi}}{a^6} \nonumber \\
\dot k &=& \frac{Q^2}{2} 
\frac{e^{-2\phi}}{b^6} \nonumber \\
\ddot \phi &=&3 h^2 + 3k^2 - \dot \phi^2 \\
3h^2 + 3k^2 &=&\dot \phi^2 
+ 2\Lambda_7 e^{-2\phi} - \frac{1}{2} e^{-2\phi} 
(\frac{P^2}{a^6} + \frac{Q^2}{b^6}) \nonumber
\end{eqnarray}
where $h$ and $k$ are logarithmic 
time derivatives of $a$ and $b$, 
respectively. If the 
cosmological term were absent\footnote{Solutions for the case
$\Lambda_7 \ne 0$ can be found by the application of 
the technique exhibited in \cite{Kal} to the solutions 
to be presented here.},
it is not difficult to see that 
if one, or both, of the charges vanish,
the equations of motion are readily 
integrable. If both charges vanish, 
we can rewrite the
dynamics in terms of normal 
coordinates as a set of free fields \cite{VG}.
When only one of the
charges vanishes the equations 
are equivalent to a Hamiltonian 
system consisting of a 
Liouville theory, or the $SU(2)$ Toda model, 
and several free fields \cite{CLW}. Moreover,
the case with a single charge 
can be obtained from a chargeless solution
by an electric-magnetic 
$SL(2,R)$ duality transformation of the reduced 
4D action (which in this case is an 
exact symmetry of the classical 4D action
because all the $U(1)$ 1-form 
fields vanish identically) \cite{CLW}. On the contrary, 
this transformation in its conventional form
cannot be employed to generate 
solutions with two charges,
because when we reduce the 
theory to 4D, assuming that the internal 
three-torus is carrying the 
nonvanishing charge to obtain the
other charge from the space-time 
axion, we can see that there arise moduli
terms which break the $SL(2,R)$ 
symmetry. Let us show this explicitly.
As the solution above has six 
translational Killing symmetries ($x$ and
$y$ translations), we could reduce the solution
down to $4D$ by eliminating 
three of the cyclic coordinates. 
In usual cases, 
after tedious but straightforward algebraic
manipulations with the reduced action, we would
have rewritten it as:
\begin{equation}
\label{fact}
S = \int d^4x \sqrt{{g_4}} e^{- \Phi_4} \Bigl\{{R_4}
 + ({\nabla} \Phi_4)^2 + \frac{1}{8} Tr \bigl(l
 \nabla m\bigr)^2
- \frac{1}{12} {H_4}^2_{\mu\nu\lambda}\Bigr\}
\end{equation}
\noindent where the subscript 
represents the $4D$ equivalents of the original 
$7D$ fields, obtained via the
Kaluza-Klein dimensional reduction.
This correctly describes our case with both
charges vanishing, or the case when only one of the charges
is zero, and we consider 
the subspace it would have wrapped around as internal.
The $\sigma$-model fields $m$ appear after
rearranging the scalar moduli fields.
The correspondence is given by
\begin{equation}\label{w4}
m~=~\pmatrix{
g^{-1}&-g^{-1}b \cr
bg^{-1}&g - b g^{-1}b \cr} ~~~~~~~~~~
l~=~\pmatrix{~0~&{\bf 1} \cr
{\bf 1}&~0~\cr}
\end{equation}
where $~g~$, $~b~$ and ${\bf 1}$ 
are $~3 \times 3~$ matrices
defined by the internal space 
components of the metric and the
axion:$~g~=~\bigl(g_{AB}\bigr)$ 
and $b~=~\bigl(b_{AB}\bigr)$.
The axion field strength can be rewritten as
\begin{equation} \label{axfsin}
 {H_4}_{\mu\nu\lambda} = 
\partial_{\mu} {B_4}_{\nu\lambda}
+ cyclic ~permutations
\end{equation}
We have assumed here that 
all the cross terms (1-forms)
between the internal space 
and $4D$ space-time vanish.
To see how the $S$-duality comes into
play, we need to first conformally rescale
the metric to the $4D$ Einstein frame, which 
is given by $\hat g_{4\mu\nu} 
= \exp(-\Phi_4) g_{4\mu\nu}$, and
then replace the three-form axion field strength in
four dimensions by its dual pseudoscalar
field. The correspondence between them is
\begin{equation}
\label{axiondef}
H_{4\mu\nu\lambda} = e^{2 \Phi_4} \sqrt{\hat g_4}
\epsilon_{\mu\nu\lambda\rho} 
\hat \nabla^{\rho} {\cal B}
\end{equation}
The hat here denotes the Einstein conformal frame.
If we introduce the complex axidilaton field
${\cal A} = {\cal B} + i \exp(-\Phi_4)$ 
we can rewrite the action (\ref{fact}) as
\begin{equation}
\label{axidilacti}
S = \int d^4x \sqrt{\hat g_4} \Bigl\{ \hat R_4
+ \frac{2 \hat \nabla_{\mu} {\cal A} \hat \nabla^{\mu}
 {\cal A}^{\dagger}}{({\cal A} - {\cal A}^{\dagger})^2}
+ \frac{1}{8} Tr \bigl(l \hat \nabla m\bigr)^2
 \Bigr\}
\end{equation}
where $\dagger$ denotes complex conjugation.
A careful examination of the equations of motion
derived from this action shows that they are
invariant under the 
following set of transformations:
\begin{eqnarray}
\label{sl2r}
{\cal A} \rightarrow {\cal A}' =
\frac{c {\cal A} + d}{a {\cal A} + b} ~~~~~~
~~~~~ cb - da =1
\end{eqnarray}
where $a, b, c, d$ are all real
numbers. 
The action also remains invariant under
(\ref{sl2r}) because the $U(1)$ gauge fields are absent.
The transformations (\ref{sl2r}) combine to form the 
$SL(2,R)$ group. This symmetry group is referred to as the
electric-magnetic $S$-duality. It is clear, however, that
the application of this symmetry doesn't offer any special
advantage when we reduce on the subspace which doesn't carry
the charge. It is not difficult to see that in this case,
the solution is $S$-selfdual, modulo trivial rescalings
(in direct analogy to, for example, dyonic black holes
in string gravity).
Thus, in order to see if we can 
use the $SL(2,R)$ group to generate
a solution with two charges, we 
have to carry out dimensional
reduction along a different path. 
Namely, to be able to mix the
axion and the dilaton in four 
dimensions in a nontrivial manner, 
we should treat the chargeless 
spatial section as a part of the
$4D$ space-time, and the charged 
spatial section as internal.
Let $y$ coordinates parametrize 
the internal space.
Because the corresponding $2$-form 
axion field formally depends on
an internal coordinate, 
say $y^3$, via $B = Q y^3 dy^1 dy^2$
(note that it is still 
invariant under translations 
$y^3 \rightarrow y^3 + \zeta$, 
since its variation
is proportional to a gauge transformation:
$\delta B = \zeta dy^1 dy^2 = d(\zeta y^1 dy^2)$) we have to
carry out the dimensional 
reduction more carefully. We cannot 
group the internal components 
of the $B$ field in the 
$4D$ $\sigma$-model form - the
internal charge $Q$ gives 
rise to an effective dilaton potential
in $4D$ instead. After 
transforming to the Einstein frame
and replacing the $4D$ 
$3$-form field with its Hodge dual as in (\ref{axiondef}),
we can write the resulting action as 
\begin{equation}
\label{axidilacti2}
S = \int d^4x \sqrt{\hat g_4} \Bigl\{ \hat R_4
+ \frac{2 \hat \nabla_{\mu} {\cal A} \hat \nabla^{\mu}
 {\cal A}^{\dagger}}{({\cal A} - {\cal A}^{\dagger})^2}
+ \frac{1}{8} Tr \bigl(l \hat \nabla m\bigr)^2
- \frac{Q^2}{2} \frac{e^{\Phi_4}}{b^6}
 \Bigr\}
\end{equation}
The last term obviously breaks the
$SL(2,R)$ invariance 
(the $\sigma$-model field
$m$ actually contains only the 
scale factor $b$; the reason that we have
chosen to keep the action in the 
above form is to underline that only the
last term breaks the $SL(2,R)$-duality)
\footnote{To get a more intuitive picture how this
term arises in four dimensions, we could have split
the original $7D$ 3-form into two pieces, one living in
the effective $4D$ internal space and the other in the
spacelike sections of the $4D$ space-time. The internal
space 3-form could then have been Hodge-dualized with
$7D$ Levi-Civita tensor, to a 4-form, now living entirely
in the $4D$ space-time. Upon the subsequent dimensional 
reduction to four dimensions, this term could have been
Hodge-dualized again, now with $4D$ Levi-Civita tensor, to
a cosmological-like term - i.e. precisely the potential
we get above.}.

It is then a little
surprising to find that this case is still
integrable. We can see this 
as follows: first, we manipulate the equations
of motion to rewrite the 
dilaton equation as 
$\ddot \phi = - \dot h - \dot k$,
and integrate this twice 
to obtain $\phi = \phi_0 + Ct - \ln ab$ ($\phi_0$ and
$C$ are the two ensuing 
integration constants). Next, we note
that if we define new 
variables $\psi$ and $\chi$ according to
\begin{equation}
\label{def}
a^3 = \frac{P}{C} 
\exp(\frac{3}{2} \chi - 3Ct - \phi_0)~~~~~~~~~~~~~
b^3 = \frac{Q}{C} 
\exp(\frac{3}{2} \psi - 3Ct - \phi_0)
\end{equation}
\noindent and we employ the scaled 
time variable 
\footnote{Note that
when $C=0$ these transformations break down. 
Nevertheless, the system is still equivalent
to an $SU(3)$ Toda molecule, as 
can be seen by the substitution
$a^3 = P\exp(3\chi/2 - \phi_0)$, 
$b^3 = Q\exp(3\psi/2 - \phi_0)$.} $\tau = Ct$, we can
rewrite the independent equations of motion as 
\begin{eqnarray}
\label{toda}
\ddot \chi &=& e^{\psi - 2 \chi} ~~~~~~~~~
\ddot \psi = e^{\chi - 2 \psi} \nonumber \\
\dot \chi^2 &+& \dot \psi^2 - \dot \chi \dot 
\psi = 6 - e^{\psi - 2 \chi} - e^{\chi - 2 \psi}
\end{eqnarray}
\noindent These equations are now 
recognized as the Toda $SU(3)$ molecule; 
they can be identified with the motion 
of a particle in two dimensions, determined by the
Hamiltonian $H = \dot \chi^2 + 
\dot \psi^2 - \dot \chi \dot \psi +  
\exp(\psi - 2 \chi) + \exp(\chi - 2 \psi)$ 
with the energy $E=6$ (because of the
chosen time normalization). In order to 
solve this system, we could invoke the
theoretical machinery of Lie symmetries 
and the Lax pair construction \cite{Toda}; suffice it
to say that this leads to 
the solution of the form 
\begin{equation}
\label{todaans}
e^{\chi} = \sum^{3}_{1} A_i e^{-\lambda_i \tau} ~~~~~~~~
e^{\psi} = \sum^{3}_{1} B_i e^{\lambda_i \tau}
\end{equation}
\noindent where the constants $A_i, B_i$ 
and $\lambda_i$ are determined by the
initial conditions. For our purposes, 
it is sufficient to use this as another ans\"atz,
simply substitute it in the equations of 
motion (\ref{toda}) and 
use the initial conditions
$\exp(\chi_0) = \alpha_1$,
$\exp(\psi_0) = \alpha_2$, 
$\exp(\chi_0) \dot \chi_0 = \beta_1$ and
$\exp(\psi_0) \dot \psi_0 = \beta_2$ 
to determine the above parameters. 

We will merely
sketch here the main points of this 
computation. Choosing $\tau = 0$ as the instant to
define the initial conditions,  we see 
that combining them with the ans\"atz (\ref{todaans})
leads to the identities $\sum^{3}_{1} A_i = \alpha_1 $,
$\sum^{3}_{1} B_i = \alpha_2 $, 
$\sum^{3}_{1} \lambda_i A_i = -\beta_1 $ and
$\sum^{3}_{1} \lambda_i B_i = \beta_2 $. 
It is convenient to
introduce new parameters $a_k$ and 
$b_k$  according to 
\begin{equation}
\label{iter1}
a_k = (\lambda_i - \lambda_k)
(\lambda_j - \lambda_k) A_k ~~~~~~~~~~
b_k = (\lambda_i - \lambda_k)
(\lambda_j - \lambda_k) B_k  ~~~~~~~~~i \ne j \ne k
\end{equation}
\noindent For now, we will assume 
that $\lambda_i \ne \lambda_j, i \ne j$. 
We will obtain the
degenerate case later,
as a limit of the nondegenerate one. 
Treating, for example, 
$\lambda_i$, $a_3$ and $b_3$ as
independent parameters we can express  the 
remaining four parameters in terms of them. This
split however is completely arbitrary and 
hence permuting the indices we obtain the general
formulae
\begin{equation}
\label{iter2}
a_k - a_j = (\lambda_j - \lambda_k) 
(\lambda_i \alpha_1 + \beta_1) ~~~~~~~~
b_k - b_j = (\lambda_j - \lambda_k) 
(\lambda_i \alpha_2 - \beta_2)
\end{equation}
Substituting these identities into the 
Toda molecule equations of motion gives
further constraints, which relate $a_k$'s and
$b_k$'s: 
\begin{equation}
\label{iter3}
a_j b_j = 1 ~~~~~~~~ \prod^{3}_{1} a_i 
= \prod^{3}_{1} b_i = -1 
~~~~~~~~ \sum^{3}_{1} \lambda_i = 0
\end{equation}
We can now combine the constraints 
(\ref{iter2}) and (\ref{iter3}) to determine
$a_k$ and $b_k$; solving these equations yields
\begin{equation}
\label{iter4}
a_j = \frac{\lambda_j \alpha_2 - \beta_2}
{\lambda_j \alpha_1 + \beta_1} ~~~~~~~~ 
b_j = \frac{\lambda_j \alpha_1 + \beta_1}
{\lambda_j \alpha_2 - \beta_2}
\end{equation}

The final step is to determine the 
``eigenvalues" $\lambda_j$. It is straightforward
to see that they are given as the 
solutions of the system of equations
\begin{eqnarray}
\label{iter5}
\sum^{3}_{1} \lambda_i &=& 0 \nonumber \\
\prod^{3}_{1} (\lambda_i \alpha_1 
+ \beta_1) &=& - (\alpha_2 \beta_1 + \alpha_1 \beta_2)  \\
\prod^{3}_{1} (\lambda_i \alpha_2 - \beta_2)  
&=& (\alpha_2 \beta_1 + \alpha_1 \beta_2) \nonumber 
\end{eqnarray}
\noindent A more transparent notation can be achieved if we 
introduce the three parameters 
$\kappa_n = \frac{1}{n} 
\sum^{3}_{1} (\lambda_i)^n, ~~~ n = 1, \ldots, 3$
(note that $\kappa_1 = 0$). With the help of these, we can see 
that the $\lambda_i$ are 
the three roots of the secular polynomial
\begin{equation}
\label{secular}
\lambda^3 - \kappa_2 \lambda - \kappa_3 = 0
\end{equation}
\noindent What remains to determine is 
the relationship between the parameters $\kappa_i$ and
the initial conditions, in order to 
specify their values. 
This we find 
from the last two constraints of (\ref{iter5}). The result
is
\begin{eqnarray}
\label{invts}
\kappa_2 &=& \frac{\beta^2_1}{\alpha^2_1} 
+ \frac{\beta^2_2}{\alpha^2_2} 
- \frac{\beta_1\beta_2}{\alpha_1\alpha_2} 
+ \frac{\alpha_1}{\alpha^2_2} 
+ \frac{\alpha_2}{\alpha^2_1} \nonumber \\
\kappa_3 &=& \frac{\beta_1\beta_2}{\alpha^2_1\alpha^2_2} 
(\beta_2\alpha_1 - \beta_1\alpha_2) 
+ \frac{\beta_1}{\alpha^2_2} - \frac{\beta_2}{\alpha^2_1} 
\end{eqnarray}
\noindent We recognize the first of these 
two equations as $\kappa_2 = H =6$ - i.e. this is
nothing else but the Hamiltonian constraint 
in a different guise. The second equation can be
rewritten as 
$\kappa_3 = \dot \chi \dot \psi (\dot \psi - \dot \chi) 
+ \dot \chi \exp(\chi - 2\psi)
- \dot \psi \exp(\psi - 2 \chi)$; this is also 
an integral of motion as can be verified by
differentiating it with respect to time, and 
using the equations of motion. It is a consequence
of a hidden symmetry of the Toda system, and 
effectively the reason why the $SU(3)$ Toda 
molecule is integrable. 

We can therefore summarize the solution as follows:
\begin{eqnarray}
\label{sol1}
e^{\chi}&=& \sum^{3}_{k=1} \frac{1}{\Delta(\lambda_k)} 
\frac{\lambda_k \alpha_2 - \beta_2}
{\lambda_k \alpha_1 + \beta_1} 
~e^{-\lambda_k \tau} \nonumber \\
e^{\psi}&=& \sum^{3}_{k=1} \frac{1}{\Delta(\lambda_k)} 
\frac{\lambda_k \alpha_1 + \beta_1}
{\lambda_k \alpha_2 - \beta_2}
~e^{\lambda_k \tau} \\
\Delta(\lambda_k) &=& 
\sum^{3}_{j=1} \prod^{3}_{l=1,l\ne j} 
(\lambda_k - \lambda_l) 
\nonumber 
\end{eqnarray}
\noindent 
Note that a simultaneous change of  
sign of the roots $\lambda_k$ and a
permution of the fields $\chi$ and 
$\psi$ amounts to time reversal. This will
be useful below.
At this point we 
could substitute this solution back 
into the ans\"atz for the 
graviton multiplet (\ref{dans}). 
A more transparent representation, in our 
opinion, can be obtained by a firsthand 
qualitative analysis of (\ref{sol1}),
which would highlight the physically 
important aspects of an Universe
described by (\ref{dans}) and (\ref{sol1}). 
One should note that
whereas the solution (\ref{sol1})
is valid for all real $\tau$, the definition
(\ref{dans}) of $\chi$ and $\psi$ implies
that the times for which the scale factors
vanish may be singularities, and we have
to examine them carefully.
Let us first eliminate unphysical parameters 
from the problem. It is easy to see
that the effect of the initial 
conditions $\alpha_1$ and $\alpha_2$ 
on the bulk dynamics is ignorable;
by the coordinate transformations 
$dx^k \rightarrow dx^k /P^{1/3}\alpha_1^{1/2}$,
$dy^k \rightarrow dy^k /Q^{1/3}\alpha_2^{1/2}$ 
we can set the charges
$P,Q$ equal to $\alpha_1^{-3/2}, \alpha_2^{-3/2}$, 
respectively. Therefore,
the effect of the initial conditions for 
$\chi$ and $\psi$ is only to set
a possible deficit angle if 
the spatial sections are
three-torii. In what follows 
we will therefore set $\alpha_1=\alpha_2=1$.

Now we can continue with the 
discussion of the solutions. There
are three subfamilies described 
by (\ref{sol1}), depending on the roots $\lambda_k$
of the secular polynomial. 
The possibilities are $i)$  roots all real and different;
$ii)$  roots all real, two of them 
degenerate; $iii)$ one root real, 
two complex (conjugates
of each other). The roots cannot all 
be imaginary because of the positivity of the
Hamiltonian constraint.  
One could expect now that the functions 
$\exp(\chi)$ and $\exp(\psi)$ are always
positive, in analogy to the case with 
a single charge. This is indeed true in
cases $i)$ and $ii)$; the Universes
described by these two subclasses of solutions
are still singular, undergoing first 
a long period of contraction until
the volumes become small enough for 
the winding mode source to overturn the
contraction of at least one (but not necessarily both) 
of the scale factors into a stage of unbounded 
expansion. Depending on the sign of the
integration constant $C$, these solutions
have either a past ($C<0$, $(-)$ branch) or a future
singularity ($C>0$, (+) branch). The case $iii)$, 
however, has a completely different singularity 
structure, where the 
spatial sections always shrink to a point 
in spite of the charges they carry. This is without an
analogue in the simplest models of Pre-Big-Bang 
cosmology.

We now present the analysis case-by-case. 
In case $i)$ we will first demonstrate
that the functions $\exp(\chi)$ and 
$\exp(\psi)$ are positive, as claimed
above.
When all the roots are real and different, 
we can order them as a decreasing
sequence; so 
suppose $\lambda_1 > \lambda_2 >\lambda_3$.
For simplicity, we can assume $\kappa_3 \ge 0$. 
This implies that $\lambda_1 >0$
and $\lambda_2, \lambda_3 \le 0$ (because 
$\kappa_3 = \prod \lambda_k$). The other
case, $\kappa_3<0$, can be reduced to this one by a
time inversion (i.e. the sign change
of $\lambda_k$'s and the interchange of $\chi$ 
and $\psi$). So we have 
$\lambda_1 >0 \ge \lambda_2 >\lambda_3$. If 
$\beta_1 + \beta_2 \ne 0$, the Eqs. (\ref{iter2}) 
tell us that $a_k \ne a_j$ for $k \ne j$. 
For suppose the opposite; if any two $a_k$'s
were equal for different $k$'s,
we would have obtained the identity 
$(\beta_1 + \beta_2)(\lambda_k - \lambda_j)=0$ 
for the pair of indices $k,j$ under
scrutiny; this is a contradiction as the 
roots are nondegenerate and
we have assumed $\beta_1 + \beta_2 \ne 0$. 
Further, $a_1 a_2 a_3 = -1$ tells us that
two of the $a_k$'s are positive and one is 
negative. We now need to show that
$a_1, a_3>0$ and $a_2 <0$; this will 
guarantee that $A_k$'s and $B_k$'s, and hence
$\exp(\chi)$ and $\exp(\psi)$, are 
always positive. We will do this by deriving
inequalities for the linear functions 
$\lambda_k + \beta_1$ and  $\lambda_k - \beta_2$.
There are four special cases to look at: 
${\bf a)}$ $\beta_1, \beta_2 >0$ ~${\bf b)}$ 
$\beta_1, \beta_2 >0$ ~${\bf c)}$ $\beta_1 >0, \beta_2<0$
with $\beta_1 + \beta_2 > 0$ and ~${\bf d)}$ 
$\beta_1<0, \beta_2 >0$ with $\beta_1 + \beta_2 > 0$.
The options with the opposite 
sign of $\beta_1 + \beta_2$ are equivalent to the
last two cases after the interchange 
of $\chi$ and $\psi$ 
(which implies the interchange
of $\beta_1$ and $\beta_2$ and hence the 
sign flip of $\beta_1 + \beta_2 $).
So we find:

${\bf a)}$ by our assumptions, 
$\lambda_1 + \beta_1$, $\beta_2 - \lambda_2$, and 
$\beta_2 - \lambda_3$ are all positive;
from $\prod^{3}_{k=1} (\lambda_k + \beta_1) 
= \prod^{3}_{k=1} (\beta_2 - \lambda_k) 
= - \beta_1 - \beta_2 <0$ and $\lambda_2 > \lambda_3$
we see that $\lambda_2 + \beta_1$ must also be positive, while 
$\lambda_3 + \beta_1$ and $\beta_2 - \lambda_1$ must be 
negative; this follows by the negativity of 
the product of these functions, and the fact that
$\lambda_3 + \beta_1>0$ and $\lambda_2 > \lambda_3$ 
would imply $\lambda_2 + \beta_1>0$, contradicting the 
assumption that their product is negative;
using these inequalities in the 
definitions of $a_k$ we immediately see that
$a_1, a_3 >0$ and $a_2 >0$; in conjunction with 
$\Delta(\lambda_1), \Delta(\lambda_3)>0$ 
and $\Delta(\lambda_2)<0$,  we find
$A_k, B_k>0$ as claimed.

${\bf b)}$ the products satisfy 
$\prod^{3}_{k=1} (\lambda_k + \beta_1) 
= \prod^{3}_{k=1} (\beta_2 - \lambda_k) 
= - \beta_1 - \beta_2>0$; 
by assumptions, 
$\lambda_2 + \beta_1$, $\lambda_3 + \beta_1$
and $\beta_2 - \lambda_1$, must 
be negative; this and $\lambda_2>\lambda_3$
imply that $\lambda_1 + \beta_1$ and 
$\beta_2 - \lambda_3$ are positive,
whereas $\beta_2 - \lambda_2$ is negative;
for suppose that $\beta_2 - \lambda_2$ 
is positive - this would imply 
that $\beta_2 - \lambda_3>0$, 
contradicting the assumption that the product
of these two expressions 
is negative; again, this means that
$a_1, a_3 >0$ and $a_2 >0$ and so $A_k, B_k>0$.

${\bf c)}$ this time, $\beta_1>0$, 
$\beta_2<0$ and $\beta_1+\beta_2>0$; thus
$\prod^{3}_{k=1} (\lambda_k + \beta_1) 
= \prod^{3}_{k=1} (\beta_2 - \lambda_k) 
= - \beta_1 - \beta_2 < 0$; since 
$\beta_1$ and $\beta_1+\beta_2$ 
are the same as in $a)$ above, we have 
$\lambda_1+\beta_1>0$, 
$\lambda_2+\beta_1>0$ and $\lambda_3+\beta_1<0$;
by the assumption about signs, 
$\beta_2 - \lambda_1<0$; therefore we
see that either $\beta_2 - \lambda_2>0$, 
$\beta_2 - \lambda_3>0$
or  $\beta_2 - \lambda_2<0$, 
$\beta_2 - \lambda_3<0$; the latter
situation would imply $a_1, a_2>0$ 
and $a_3<0$; but by 
$a_3-a_2=(\lambda_2-\lambda_3)(\lambda_1+\beta_1)$ 
(see Eq. (\ref{iter2}))
we would get $a_3>a_2$, 
contradicting the latter case; hence
it must be  $\beta_2 - \lambda_2>0$, 
$\beta_2 - \lambda_3>0$ and
so $a_1, a_3>0$ and $a_2>0$ 
again yielding $A_k, B_k>0$.

${\bf d)}$ finally, $\beta_1<0$, $\beta_2>0$ 
and $\beta_1+\beta_2>0$; so
$\prod^{3}_{k=1} (\lambda_k + \beta_1) 
= \prod^{3}_{k=1} (\beta_2 - \lambda_k) 
= - \beta_1 - \beta_2 < 0$; since now
$\beta_2$ and $\beta_1+\beta_2$ are 
the same as in $b)$, we get
$\beta_2-\lambda_1<0$, $\beta_2-\lambda_2<0$ 
and $\beta_2-\lambda_3>0$;
the signs imply that $\lambda_1+\beta_1>0$; 
the remaining options are
$\lambda_2+\beta_1<0$, $\lambda_3+\beta_1<0$  or
$\lambda_2+\beta_1>0$, $\lambda_3+\beta_1>0$; 
the latter case
is again contradictory, because it would 
lead to $a_1, a_2>0$ and $a_3<0$
whereas by
$a_3-a_2=(\lambda_2-\lambda_3)(\lambda_1+\beta_1)$ 
we would get $a_3>a_2$; hence 
it must be  $\lambda_2+\beta_1<0$, 
$\lambda_3+\beta_1<0$ and
so $a_1, a_3>0$ and $a_2>0$, 
once more leading to $A_k, B_k>0$.

To completely cover the case of 
real nondegenerate roots, we need to review the situation 
when $\beta_1+\beta_2=0$. Fortunately this
is simple enough to be 
evaluated directly. The equality
$\beta_1 = - \beta_2 = \beta$ and the 
Hamiltonian constraint set 
$\beta^2=4/3$. Moreover, the condition
$\prod^{3}_{k=1} (\lambda_k + \beta) = 0$ 
tells us that one
of the roots must be equal to 
$-\beta$. So we factor the cubic
secular polynomial, to find the roots:
$-\beta, \pm\sqrt{5} + \beta/2$
with $\beta=\pm 2/\sqrt{3}$. 
Ordered roots corresponding to $\beta=-2/\sqrt{3}$
are $\sqrt{5} + 1/\sqrt{3}$, $-2/\sqrt{3}$ 
and $-\sqrt{5}+ 1/\sqrt{3}$ 
(the $\beta$ of the opposite 
sign is equivalent to this one by a permutation
of our degrees of freedom, as 
in previous cases). Clearly
$a_1=a_3=1$, but because $\lambda_2+\beta=0$,
the formula for $a_2$ is 
ill-defined, leading to an expression of the
form $\propto 0/0$. A more 
careful calculation should employ
$a_2-a_3=(\lambda_3-\lambda_2)
(\lambda_1+\beta)=-2$ to give 
$a_2=-1$. Lastly, using the 
values for $\lambda_k$ and the formulae for $A_k$, $B_k$, we
find $A_1 = B_1 = 
1/2\sqrt{5}(\sqrt{5}+\sqrt{3})$, 
$A_2 = B_2 = 1/2$ and  
$A_3 = B_3 = 1/2\sqrt{5}(\sqrt{5}-\sqrt{3})$ -
i.e. $A_k, B_k>0$. 

In sum, we see that when the roots are real 
and nondegenerate, the
functions $\exp(\chi)$ and 
$\exp(\psi)$ remain positive, each 
being a linear combination of three 
positive terms. This is reminiscent of
the case of a 
single charge and a constant internal
space radius \cite{CLW}, whereby we might be tempted to
conjecture that the spatial sections
of this solution can never shrink to 
a point. Intuitively this would agree with
the assertion that a winding 
mode causes an initially contracting
scale factor of a subspace around 
which the mode wraps to bounce
into expansion. The analogy is misleading, however,
because the scale factors are not identical 
with the solution of the Toda equations and
we need 
to carry out a more careful inspection of the
solutions in order to see what happens. 

First of all, the solutions are
still singular,  as
can be seen from the inspection of 
the Ricci scalar. A straightforward computation yields
\begin{equation}
\label{ricci}
R = 6e^{2\phi}\Bigl(\dot h + \dot k 
+ \dot \phi (h + k) + 2h^2 + 2k^2 + 3hk \Bigr)
\end{equation}
Let us look at this expression 
for the case of three real nondegenerate
roots when $\tau \rightarrow \pm \infty$.
As before, $\lambda_1>0 \ge \lambda_2>\lambda_3$.
So in the limits we have 
\begin{eqnarray}
&& e^{\chi} \rightarrow \frac{a_3}{\Delta(\lambda_3)}
e^{-\lambda_3 \tau}
~~~~~~~~e^{\psi} \rightarrow \frac{1}
{a_1 \Delta(\lambda_1)} 
e^{\lambda_1 \tau} ~~~~~~~~ 
{\rm as~~} \tau \rightarrow \infty
\nonumber \\
&& e^{\chi} \rightarrow \frac{a_1}{\Delta(\lambda_1)}
e^{-\lambda_1 \tau}
~~~~~~~~e^{\psi} \rightarrow \frac{1}
{a_3 \Delta(\lambda_3)}
e^{\lambda_3 \tau} ~~~~~~~~ 
{\rm as~~} \tau \rightarrow -\infty
\label{limits}
\end{eqnarray}
The expression for $\phi$ in terms of 
the scale factors $a$ and $b$ allows us
to rewrite the Ricci scalar as follows:
\begin{equation}
\label{ricspec}
R=6C^2 e^{2\phi_0} e^{6\tau - \chi - \psi}
\Bigl(\frac{\ddot \chi + \ddot \psi}{2} + 
\frac{\dot \chi^2 + \dot \psi^2 
- \dot \chi \dot \psi}{2} 
- \ddot \chi - \ddot \psi \Bigr)
\end{equation}
The limits (\ref{limits}) imply 
$\dot \chi \rightarrow -\lambda_3$,
$\dot \psi \rightarrow \lambda_1$
when $\tau \rightarrow \infty$,
and $\dot \chi \rightarrow -\lambda_1$,
$\dot \psi \rightarrow \lambda_3$ when 
$\tau \rightarrow -\infty$, with 
$\ddot \chi, \ddot \psi \rightarrow 0$
in both cases. The Ricci curvature 
therefore behaves as
$R \propto \exp\bigl((6 \mp (\lambda_1 
- \lambda_3))\tau \bigr)$
as $\tau \rightarrow \pm \infty$, 
while remaining finite for all other
times. Using the constraints for $\kappa_1$ and $\kappa_2$
(Eqs. (\ref{iter5})-(\ref{invts})),
we can show that 
$6 \mp (\lambda_1 - \lambda_3) > 6 - \sqrt{24} >0$.
Thus, as $\tau \rightarrow \infty$, the curvature
diverges, and as $\tau \rightarrow -\infty$, the
curvature vanishes\footnote{The same is true
for the special case $C=0$ mentioned 
in footnote 2.}. 
Further, we see from the expressions for the
Hubble parameters that in the limit $\tau \rightarrow -\infty$
both scale factors are decreasing - i.e. the spatial sections
are collapsing. This follows from the limits of the logarithmic 
derivatives of scale factors, which can be evaluated from 
(\ref{def}) and (\ref{limits}): 
$h \rightarrow -(\lambda_1 + 2)/2$
and $k \rightarrow -(|\lambda_3| + 2)/2$.
The limit $\tau \rightarrow \infty$ is more ambiguous;
to study it, let us substitute the solutions
and the choice of the lapse function back into the
ans\"atz for the metric. In the limit
$\tau \rightarrow \infty$ we find (recall that $\lambda_3 <0$)
\begin{equation}
\label{metric}
ds^2 \rightarrow 
- K_{\tau} e^{(\lambda_1 - \lambda_3 - 6)\tau} d\tau^2
+  K_{x} e^{-(\lambda_3 + 2)\tau} d\vec x^2 
+  K_{y} e^{(\lambda_1 - 2)\tau} d\vec y^2 
\end{equation}
where $K_{\tau},  K_{x},  K_{y}$ are numerical constants 
irrelevant for the discussion that follows. As we have commented 
above, $6 - \lambda_1 + \lambda_3 >0$, and so the appropriate
comoving time is $T= \exp((\lambda_1 - \lambda_3 -6)\tau/2)$, 
and in the limit $\tau \rightarrow \infty$, $T \rightarrow 0$.
In terms of it, the metric is
\begin{equation}
\label{metric1}
ds^2 \rightarrow 
- \bar K_{\tau} dT^2
+ \bar K_{x} \frac{1}{T^{n_1}} d\vec x^2 
+ \bar K_{y} \frac{1}{T^{n_2}} d\vec y^2 
\end{equation}
where $n_1 = 2(|\lambda_3|-2)/(6 - \lambda_1 - |\lambda_3|)$,
$n_2 = 2(\lambda_1-2)/(6 - \lambda_1 - |\lambda_3|)$. Clearly, 
the signs of these two numbers determine the 
qualitative behavior of the solutions: for $n_k$ positive,
the scale factor expands superexponentially, and for 
$n_k$ negative it shrinks to zero according to a simple
power law. This kind of behavior is of course familiar from the
simplest Pre-Big-Bang scenario; what is novel is the combination
of these possibilities for the given values of charges.
The allowed combinations are simultaneous pole expansion of both
spatial sections (as, for example when 
$\lambda_1 = - \lambda_3 =
\sqrt{6}$) and the contraction of one spatial section
accompanied with the pole expansion of the other. The 
simultaneous contraction of both spatial sections to a point
is impossible; this can be seen as follows.
The Hamiltonian constraint, $\kappa_2 = 6$ tells us that the
roots $\lambda_k$ live on a sphere of radius $\sqrt{12}$~:
$\sum \lambda^2_k = 12$. Using 
$\lambda_1 >0 \ge \lambda_2 > \lambda_3$ and assuming 
simultaneous contraction of both scale factors (and hence
$\lambda_1 < 2, |\lambda_3|<2$), we get 
$\sum \lambda^2_k < 3 \times 4 = 12$ - which is impossible.
The last ingredient of the phase-space description 
of evolution is the behavior of the coupling constant.
In the limits $\tau \rightarrow \pm \infty$,
the coupling of the theory 
$\exp(\Phi_7) = \exp(\phi) a^3 b^3$
behaves as $\exp(\Phi_7) = 
\exp(\chi + \psi - 3 \tau) \rightarrow 
\exp((\pm(\lambda_1- \lambda_3) - 3) \tau)$. From this we see
that in the limit $\tau \rightarrow -\infty$ the coupling 
diverges for all values of $\lambda_k$, while 
for $\tau \rightarrow \infty$ the coupling 
diverges when $\lambda_1 + |\lambda_3| >3$ (all the cases 
of simultaneous expansion of both scale factors, and 
some of the cases with one scale factor increasing 
and the other decreasing), and vanishes when 
 $\lambda_1 + |\lambda_3| <3$ (the remaining cases when 
one spatial section asymptotically superinflates and the
other shrinks to a point).

With these details, a coherent picture
emerges at last: the phase-space of the problem can be
divided into three different sectors, distinguished by
the behavior of the solution near the
singularity. In one sector, as the singularity is 
approached, the evolution is in the strong coupling
regime with both scale factors growing beyond limit.
In the other sector, the coupling is still strong, but
instead of simultaneous expansion, one of the spatial
sections shrinks to a point. In the last sector of the phase
space again one of the scale factors grows 
and the other decreases, but the system evolves towards
the weak coupling regime.
All the solutions are smoothly connected to their other
limit, where they approach an almost flat space where
both spatial sections contract simultaneously, and 
the string coupling is in the strong coupling regime.
By time inversion (reversing the 
sign of $C$), we see that both directions
of evolution are admissible: we either start from one of the
singular regions and evolve towards an asymptotically flat
space ($(-)$ branch solutions) or begin from a strongly coupled
asymptotically flat space and progress towards one of the 
singular sectors ((+) branch solutions).
This  generalizes
the simplest versions of 
the Pre-Big-Bang scenario. 
The difference is that,
depending on the initial conditions, the 
solutions studied so far are qualitatively similar to both the
chargeless (solutions evolving throughout the 
allowed range of variation of the string coupling) 
and the single charge case (solutions whose dynamics 
is confined to the strong coupling regime).
The bottom line is 
that the singularity cannot be
removed at the classical level 
from any solution characterized by
a set of real nondegenerate roots. 

The case with real degenerate roots is qualitatively
similar to the previous case. We will here prove only 
that the functions $\exp(\chi)$ and $\exp(\psi)$ never
vanish for any finite value of $\tau$. 
This is again the essential property of
the solution - the rest follows straightforwardly. 
We will view this case as a
limiting situation of the previous one. In order to quantify 
this, let us introduce a parameter 
$\epsilon = \lambda_2 - \lambda_3$, rewrite the
solution (\ref{sol1}) in terms of it, and take the
limit $\epsilon \rightarrow 0^+$. To do this, note that
the roots $\lambda_k$ can be rewritten as 
$\lambda_1 = \lambda - \epsilon$, 
$\lambda_2 = \epsilon - \lambda/2$ and $\lambda_3 = - \lambda/2$
($\lambda$ is the nondegenerate root of the secular equation).
Then we can write
\begin{eqnarray}
\label{expans}
e^{\chi} &=& \sum^{3}_{k=1} 
\frac{a_k(\epsilon)}{\Delta_k(\epsilon)} 
~e^{-\lambda_k(\epsilon) \tau} \nonumber \\
e^{\psi} &=& \sum^{3}_{k=1} 
\frac{1}{a_k(\epsilon) \Delta_k(\epsilon)} 
~e^{-\lambda_k(\epsilon) \tau}
\end{eqnarray}
where $\Delta_k$ are at most quadratic polynomials in
$\epsilon$, as can be seen from their definition in 
(\ref{sol1}). Also, by their definition, the coefficients
$a_k$ are linear fractions in $\epsilon$. Hence the
solution, being a finite linear combination of a 
finite product of functions analytic in $\epsilon$ must
also be analytic in it; so we can take the limit 
$\epsilon \rightarrow 0^+$, which is unique and reproduces the
case of two degenerate roots. It is then easy to verify that the
result of taking the limit can be written as
\begin{eqnarray}
\label{expans1}
e^{\chi} &=& \frac{a_1(0)}{\Delta_1(0)} 
~e^{-\lambda \tau} 
+ p(\tau) ~e^{\lambda \tau /2} \nonumber \\
e^{\psi} &=& \frac{1}{a_1(0)\Delta_1(0)} 
~e^{\lambda \tau} 
+ q(\tau) ~e^{-\lambda \tau /2}
\end{eqnarray}
where $p(\tau)$ and $q(\tau)$ are some linear functions
of $\tau$, $p(\tau)$ monotonically decreasing and $q(\tau)$
monotonically increasing. To see that $\exp(\chi)$ 
and $\exp(\psi)$ never vanish, it is sufficient to prove that
the lines $p(\tau)$ and $q(\tau)$ never intersect the
curves $\exp(-3\lambda/2 \tau)$ and $\exp(-3\lambda/2 \tau)$.
The flow of the curves is consistent with this assertion: 
the lines generally stay below the exponential curves. We
still need to exclude the possibility for any intersections, or
in fact even osculations of the lines with the exponentials. 
The following qualitative argument does precisely that. 
Consider the case when $\epsilon$ is very small but not zero.
Suppose then that any of the two functions of the solution
in the limit $\epsilon = 0$, say 
$\exp(\chi)$, vanishes for some $\tau$. Analyticity of 
$\exp(\chi(\epsilon))$ ensures that we can 
choose a sufficiently small negative $\epsilon$ such that
$\exp(\chi(\epsilon)) <0$. But this would contradict
the fact that  $\exp(\chi)$ 
and $\exp(\psi)$ are always positive for real nondegenerate 
roots, proven above. Hence we conclude that the
same must be true when two of the roots are degenerate.
The rest of the analysis for this case can now proceed along 
the same lines followed in the case of nondegenerate roots,
leading to similar conclusions.

The last case we need to consider is when two roots are 
complex (conjugates). This case is dramatically different 
from the two previous situations, as we will now show.
We can write down the roots as $\lambda_1 = \lambda$, 
$\lambda_2 = - \lambda/2 + i \mu$ and 
$\lambda_3 = - \lambda/2 - i \mu$. The solutions then become
\begin{eqnarray}
\label{complex}
e^{\chi} &=& \Bigl(\frac{a_1}{\Delta_1} 
~e^{-3\lambda \tau/2} + 2 {\rm Re}~
\frac{a_3}{\Delta_3} 
~e^{i \mu \tau} \Bigr) 
~e^{\lambda \tau /2} \nonumber \\
e^{\psi} &=& \Bigl(\frac{1}{a_1 \Delta_1} 
~e^{3\lambda \tau/2} + 2 {\rm Re}~
\frac{1}{a_3 \Delta_3} 
~e^{- i \mu \tau} \Bigr) 
~e^{-\lambda \tau /2}
\end{eqnarray}
Introducing the trigonometric notation and defining
$\rho_1 = |a_3/\Delta_3|$, $\rho_2 = 1/|a_3 \Delta_3|$,
$\delta_1 = {\rm Arg}~ a_3/\Delta_3 $ and 
$\delta_2 = - {\rm Arg}~a_3 \Delta_3$, we can rewrite these 
equations as 
\begin{eqnarray}
\label{trig}
e^{\chi} &=& \Bigl(\frac{a_1}{\Delta_1} 
~e^{-3\lambda \tau/2} + \rho_1 \cos(\mu \tau + \delta_1) \Bigr) 
~e^{\lambda \tau /2} \nonumber \\
e^{\psi} &=& \Bigl(\frac{1}{a_1 \Delta_1} 
~e^{3\lambda \tau/2} + \rho_2 \cos(\mu \tau + \delta_2)\Bigr) 
~e^{-\lambda \tau /2}
\end{eqnarray}
It should be obvious from the these expressions that 
for a sufficiently large (small) $\tau$, $\exp(\chi)$
($\exp(\psi)$) must vanish. The reason lies in the
oscillatory nature of the cosines, which for some $\tau$
can, and will, turn negative enough to cancel the
exponentials. However, all the
zeros of the expressions in (\ref{trig}) are simple. This
can be verified by looking at the function 
$f(z) = \eta \exp(z) + \zeta \cos(z)$ and computing the 
period around one of the zeros: $\Pi = (1/2i\pi) \oint df/f = 1$.
An even simpler way to see this is to plot the
cosine and the exponential against each other and compare the
tangents at the intersection points - the fact that they
are different guarantees that the roots are simple. 
As a corollary, the first derivative of 
$\chi$ must have a simple pole with unit residue 
at the location of the root of $\exp(\chi)$.
Also, we see from (\ref{trig}) that the other function
from the set of solutions can 
neither vanish nor diverge here. 
We can now reexamine the Ricci curvature for this case, and with 
the help of the equations of motion rewrite it as
\begin{equation}
\label{riccomp}
R = 6C^2 e^{2\phi_0 + 6\tau}\Bigl(\frac{e^{-3\chi}}{4}
+ \frac{e^{-3\psi}}{4} + \frac{3}{2} e^{-\chi-\psi}
+ \frac{e^{-\chi-\psi}}{2} \dot \chi \dot \psi
- e^{-\chi-\psi} \dot \chi - e^{-\chi-\psi} \dot \psi \Bigr)
\end{equation}
In the limit when $\tau$ approaches the location of the root
of $\exp(\chi)$, the term of the leading order of divergence 
is $\exp(-3\chi)$; it blows up as $1/(\tau - \tau_0)^3$, as
our simple analysis of the zeros and poles combined with
power counting shows. Hence there is a curvature 
singularity at the location of any of the roots of 
$\exp(\chi)$, and similarly, of $\exp(\psi)$.
In response, we conclude that the
case of two complex (conjugate) roots is pathological: all
solutions have both a past and a future singularity, where
one of the scale factors must vanish. We also note that this
subfamily of solutions is divided into two sectors: in one,
$\exp(\psi)$ is always nonzero, while $\exp(\chi)$ vanishes
twice in the lifetime of the Universe, and vice versa. These two
groups of solutions are divided by a single solution
straddling $\tau=0$, where the past singularity corresponds to 
the first root of $\exp(\chi)$, and the future singularity to the
first root of $\exp(\psi)$ (or vice-versa, depending on the sign of 
of $C$). Finally, the string coupling is again given by
$\exp(\Phi_7) = \exp(\chi + \psi - 3 \tau)$. When the solution
approaches the singularity, the coupling vanishes, and at 
all other times it remains finite. Thus, the solutions 
from this subfamily evolve between two consecutive zeros of 
scale factors, always remaining in the weak coupling region.

If we take any of the solutions above as a starting point
and reduce it to $4D$ such that the resulting metric is
of the Robertson-Walker form, we will see that all those 
modular cosmologies still suffer from singularities. The only
difference may be in deforming the
phase space in certain ways - for example, redefinitions
of effective $4D$ couplings may change the strength
of couplings in the phase space of the lower-dimensional theory.
Still, no curvature singularity can be removed in such
a way, as can be noted from the fact that the reduction does 
not affect the $4D$ space-time metric, and that in all the
solutions discussed above the curvature singularity 
is encoded democratically in both scale factors. Nevertheless,
the reduction to fewer dimensions may permit us to utilize the
duality symmetries for extending the solutions in nontrivial
ways. We will merely quote here the possibility of reducing
the solution on one of the spatial directions to $6D$,
and using either the $U$ duality between the IIA and 
the IIB theory, or the $SL(2,R)$ NS-NS/RR-RR duality
in IIB theory as a solution-generating technique. This would
of course give rise to Toda models carried by forms of 
different rank (e.g. a 2-form and a 1-form, as mentioned
very recently in \cite{LOW}). We will not delve into a detailed
investigation of such solutions here. Instead we will point an 
interesting relationship between our solutions,
reduced to $4D$, and a class of modular Bianchi II cosmologies.
If we look at the metric and the axion of the 
$4D$ version  of our ans\"atz, ignoring the internal
degrees of freedom, 
\begin{eqnarray}
\label{dansB}
d{s_4}^2 &=& -n(t)^2 dt^2 + a(t)^2  d\vec x^2  \nonumber \\
B_4 &=& P x^1 dx^2 \wedge dx^3  \\
\Phi_4 &=& \Phi_7 - 3 \ln b  \nonumber
\end{eqnarray}
we see that the translations along $x^2$ and $x^3$ are manifest
symmetries of the solution (the translation in the $x^1$ 
direction is also a symmetry since the change of the axion
2-form is just a gauge transformation, as discussed on page 6).
So we can transform this solution
with the simplest of all $T$-dualities,
the $R\rightarrow 1/R$ map  \cite{Busch}. We can choose to
dualize with respect to $x^2$, in which case the map is
\begin{eqnarray}
\label{Bucher}
g'_{22} &=& \frac{1}{g_{22}} ~~~~~~~~~~~
g'_{2\mu} = \frac{B_{2\mu}}{g_{22}} ~~~~~~~~~~~
B'_{2\mu} = \frac{g_{2\mu}}{g_{22}} \nonumber \\
&& g'_{\mu\nu} = g_{\mu\nu} - \frac{g_{2\mu} g_{2\nu} -
B_{2\mu} B_{2\nu}}{g_{22}} \\
&& B'_{\mu\nu} = B_{\mu\nu} - \frac{g_{2\mu} B_{2\nu} -
B_{2\mu} g_{2\nu}}{g_{22}} \nonumber
\end{eqnarray}
The dual solution can be written as
\begin{eqnarray}
\label{Bianchidu}
d{s'_4}^2 &=& -n(t)^2 dt^2 + 
a(t)^2  \bigl((dx^1)^2 + (dx^3)^2 \bigr) + 
\frac{1}{a(t)^2} \bigl(dx^2 + Px^1 dx^3 \bigr)^2  \nonumber \\
&& ~~~~~~ B'_4 = 0 ~~~~~~~~~~~~~~~~
\Phi'_4 = \Phi_4 - 2 \ln a 
\end{eqnarray}
Without engaging in a detailed analysis of this solution,
let us merely point out that it belongs to a different class 
of homogeneous, but not isotropic cosmologies: the Bianchi II
Universe, defined by the following algebra of the spatial 
basis 1-forms $e^1 = dx^1$, $e^2 = dx^2 + P x^1 dx^3$ 
and $e^3 = dx^3$:
\begin{equation}
\label{Bianchi}
de^1 = de^3 = 0 ~~~~~~~~~ de^2 = P e^1 \wedge e^3
\end{equation}
This change of the algebra of basis 1-forms reflects the
change of the topology of space-time 
under the duality (\ref{Bucher}).
We find this relationship rather peculiar. While both solutions,
in their classical form, are singular, it is known that 
the $\alpha'$ expansion resummation of certain anisotropic
cosmological models can lead to the removal of the singularity
present at the classical level, and thus to the Planck-scale 
resolution of the graceful exit problem \cite{KK,ART,EM}. 
A similar argument for
homogeneous and isotropic models is still lacking, in part 
because of the absence of a conformal field-theoretic
description needed for such an investigation. Perhaps such
an approach may be aided by a relationship of the type 
considered here. The $T$-duality, and perhaps the other 
dualities of string theory, may provide 
a natural way to introduce such anisotropies \cite{BKB}
into the consideration of apparently homogeneous and isotropic
models, and so allow for a resolution of the singularity 
even in this case. 

In conclusion, we have shown how a nontrivial 
$SU(3)$ Toda molecule dynamics arises in string cosmology 
from a gravity-mediated interaction of two winding modes.
The model is exactly integrable, and leads to a broad family
of solutions which are all singular, either in the future
or in the past, or both. Thus all solutions suffer from the
graceful exit problem, if viewed as a realization of
the Pre-Big-Bang cosmology. In addition, there are solutions
with string winding modes wrapped around spatial subspaces,
where the subspaces can shrink to zero volume. This aberration
from the behavior anticipated earlier comes about as a
consequence of the gravity-mediated interaction between the modes.
We have also pointed out a duality 
relationship of our solutions with a Bianchi II anisotropic
model, and have suggested that while the model is classically 
singular, the anisotropies may lead to its Planck-scale
regularization, after the $\alpha'$ expansion is resummed.
The relationship we have pointed out could also be viewed as an
indication that anisotropies in string cosmology may not have
as drastic consequences for the evolution of the Universe as
they do in General Relativity - for if they can be
removed altogether by a duality map, believed to relate 
equivalent physical pictures,  they need not be a physically 
relevant notion. The clarification  of the 
relationship between the singularity, anisotropies 
and duality therefore seems to warrant further 
investigation. 
\vskip 1.0truecm
{\it Note added in proof:} While this report was being prepared,
there appeared an article by A. Lukas, B. Ovrut and D. Waldram,
\cite{LOW}, which overlaps with a portion of the work 
presented here.
\vskip 1.0truecm
\noindent {\bf Acknowledgements}
\vskip 0.7truecm
Thanks are due to R.C. Myers for helpful comments.
This work was supported in part by  NSERC of Canada, in 
part by Fonds FCAR du Quebec, and
in part by an NSERC postdoctoral fellowship.

\end{document}